# Modeling the atomic structure in the vicinity of the spherical voids and calculation of void growth rate anisotropy in bcc iron and tungsten

A.V. Nazarov[1,2], A.P. Melnikov[1], A.A. Mikheev[3], I.V. Ershova[1]

[1]National Research Nuclear University MEPhI, (Moscow Engineering Physics Institute), 31, Kashirskoe shosse 115409, Moscow, Russia.

[2]Institute for Theoretical and Experimental Physics named by A.I. Alikhanov of NRC "Kurchatov Institute", 25 Bolshaya Cheremushkinskaya str.,117218, Moscow, Russia

[3] The Kosygin State University of Russia, 33 Sadovnicheskaya str., 117997, Moscow, Russia.

**E-mail:** avn46@mail.ru

**Abstract**. We study the influence of the atomic structure in the vicinity of voids on their growth rate anisotropy. In the first part, we model the atomic structure in the vicinity of nanovoids in α-Fe and W using the advanced Molecular Statics method. In the second part, we use the earlier obtained equations that taking the influence of elastic fields into account to calculate the shifting rate of the void surface elements, and to evaluate the components of the strain tensor we use the atomic structure modeling results from the first part. The calculations are performed for voids of several sizes at certain oversaturation's in a wide temperature range. The simulation results for the mentioned metals with a bcc structure show that displacements of atoms located along the crystallographic directions of the <100>, <110>, <111> types in the vicinity of the voids are significantly different, and this anisotropy of atom displacements leads to a reduction of spherical symmetry for the shifting rate of the surface elements. As the result, the initially spherical void shape becomes faceted.

## 1. Introduction

During the irradiation of materials defects such as voids and bubbles are formed, and this changes the properties and characteristics of the material. Therefore, the study of the void growth kinetics is necessary to predict material behavior under work conditions, and these issues are under special attention in the works of many authors [1-5]. It is often seen at electron microscopy images that void, initially having the spherical shape, gradually change it to faceted [3, 4, 6]. According to the works [3, 7], several factors influence the change in the void shape upon irradiation: surface energy anisotropy for different crystallographic planes, preferred adsorption of atoms on certain crystallographic surfaces and vacancy diffusion fluxes anisotropy in the vicinity voids.

Ordinarily the elastic fields around the voids are not taken into account in models that predicting the material's behavior under irradiation [1]. However, as shown previously [8,9], these fields in the case of nanovoids have a significant effect on vacancy fluxes, and, consequently, on the growth and dissolution rates of voids. According to the results of [10] the X-axis component of the vacancy flux in the zero approximation of the elastic field effect is described by the following equation:

$$J_x = -D_V \left[ \frac{\partial c}{\partial x} - c \frac{K^V}{kT} \frac{\partial Sp\varepsilon}{\partial x} \right], \tag{1}$$

where $J_x$ is the vacancy diffusion flux in the direction of X-axis, $c$ is the vacancy concentration, $D_V$ is the vacancy diffusion coefficient in an ideal system, $K^V$ is the coefficient that determines the elastic field contribution to the vacancy flux [10], $\varepsilon$ is the strain tensor, $Sp\varepsilon$ is the strain tensor trace, $k$ is the Botzmann constant, $T$ is the temperature.

Usually displacement fields in the vicinity of defects are determined by solving the isotropic theory of elasticity equations. The solution of the equations for an isolated void of a spherical shape with radius $R$ has the form [11]:

$$u_i = C_1 \frac{x_i}{r^3}, \tag{2}$$

where $u_i$ is the atom displacement from the initial position in the $x_i$ direction, $r$ is the distance from the center of the void, and $C_1$ determined by both the characteristics of the material and the void radius:

$$C_1 = -\frac{1+\nu}{2E} \gamma R^2, \tag{3}$$

where $R$ is the radius of the spherical void, $\gamma$ is surface energy, $\nu$ is the Poisson's ratio, $E$ is the Young's modulus.

The components of the strain tensor are determined by the equation:

$$\varepsilon_{ij} = \frac{1}{2}(\frac{\partial u_i}{\partial x_j} + \frac{\partial u_j}{\partial x_i}). \tag{4}$$

When differentiate the displacement (Equation (2)) we get the equation for the diagonal components of the strain tensor:

$$\varepsilon_{ii} = C_1(\frac{1}{r^3} - \frac{3x_i^2}{r^5}), \tag{5}$$

and the trace of the strain tensor is a constant, and its derivative is zero:

$$\nabla Sp\varepsilon = 0. \tag{6}$$

Thus, elastic fields do not affect diffusion fluxes and void growth rates, and in that connection voids are often called neutral sinks [1]. In our work, we use the Advanced Molecular Static method for simulation atomic structure in the vicinity of the voids and finding atom displacements [12-15]. This approach takes into account the discreteness of the structure and its anisotropy. As a result, the atoms' displacements for different crystallographic directions differ significantly, and the trace of the strain tensor is different from the constant and its gradient is not zero. Therefore, an additional term appears in the equations for vacancy fluxes, and this term differ for various crystallographic directions.

The model is described in detail in [9], there are also some results for α-Fe. The results of modeling the structure are used to calculate in the next section the dependences of the shifting rate of the void surface elements on the oversaturation in metals with a bcc structure (α-Fe and W) over a wide temperature range. Moreover, these dependencies for crystallographic directions of types <100>, <110>, <111> are significantly different, and thus, the elastic field anisotropy in the vicinity of the void can lead to a change in the shape of the initially spherical voids in the irradiated materials.

## 2. Method

Atoms surround a spherical void with radius $R$, the positions of atoms are determined by the radius vector $r$. The system is divided into two zones – main computational cell (zone I) and elastic medium in which atoms are immersed (zone II). The zone structure is shown on Fig. 1.

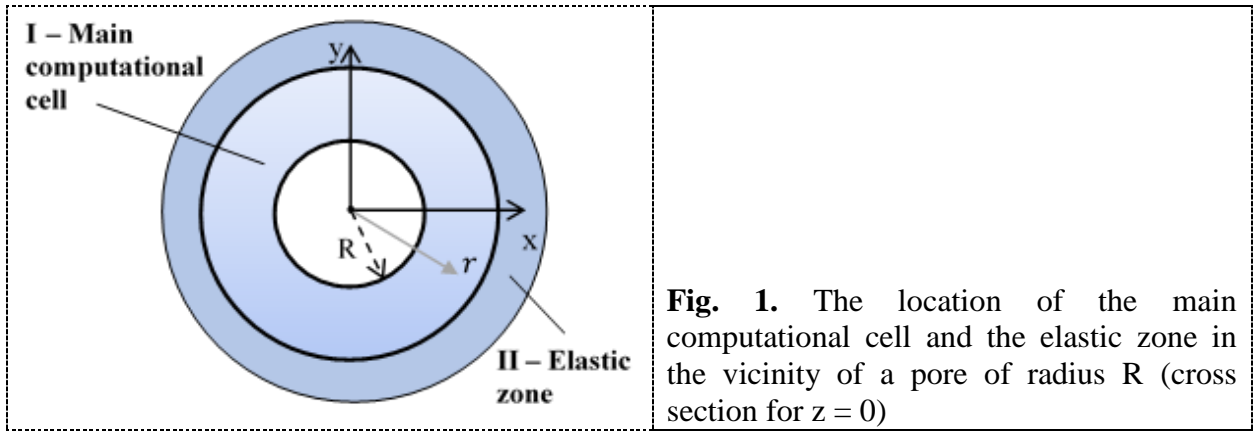

**Fig. 1.** The location of the main computational cell and the elastic zone in the vicinity of a pore of radius R (cross section for z = 0)

The atomic coordinates of the first zone are calculated using the variational procedure of the MS method. The atoms surrounding the main computational cell are located in an elastic medium, and their displacements are determined based on solutions of the elasticity theory equations:

$$u = C_1 \frac{r}{r^3}. \qquad (7)$$

An important feature of the model is a self-consistent iterative procedure for calculating the atoms' positions in the main computational cell and calculating the constant $C_1$, which determines the displacements in the elastic zone. The constant $C_1$ is calculated using equation (7) based on the results of modeling atomic displacements in the spherical layer located approximately in the middle between the defect and the boundary of the main computational cell. The simulation results indicate a stable convergence of the iterative procedure. The scheme of the iterative algorithm is presented on Fig. 2.

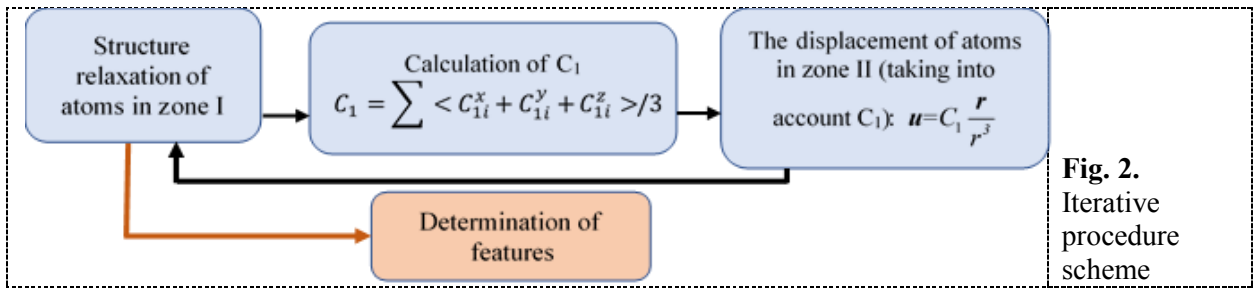

**Fig. 2.** Iterative procedure scheme

The model allows one to obtain a structure in the vicinity of voids [9,16]. The next section presents the simulation results for voids of various sizes in α-Fe and W that are further used in calculating the shifting rate of void surface elements.

## 3. Simulation results of atomic structure in vicinity of nanovoids

Figures 3 and 4 show the dependences of atomic displacements for crystallographic directions of the type <100>, <110>, <111> on the distance to the void center in α-Fe and W.

Calculations of atomic displacements based on the solution of the isotropic theory of elasticity equation are also given for comparison for voids of the same size. These solutions have spherical symmetry and are independent of angles.

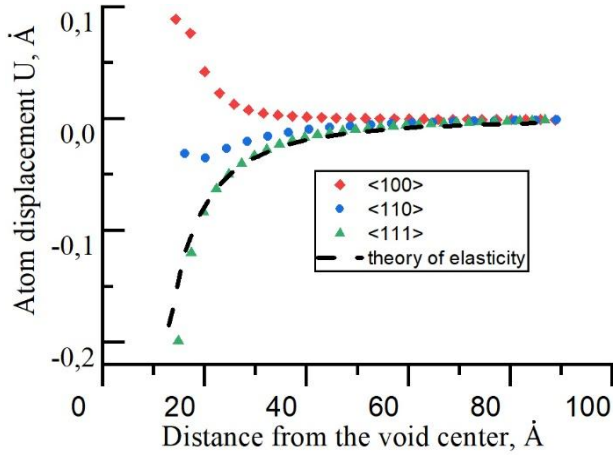 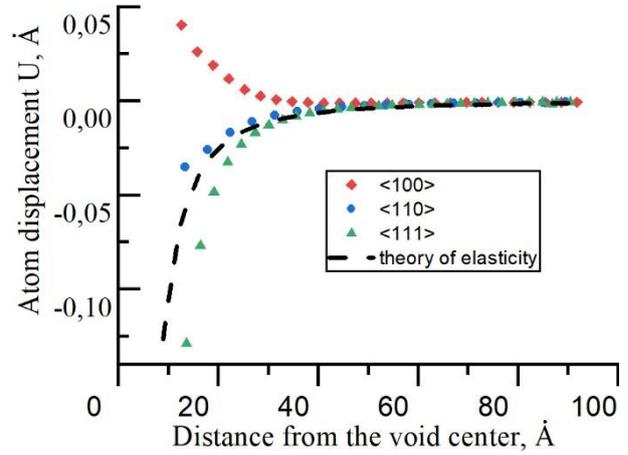

**Fig. 3.** The atoms' displacements for various directions in α-Fe for nanovoid (R = 15.14 Å)

**Fig. 4.** The atoms' displacements for various directions in W for nanovoid (R = 10.96 Å)

In contrast to the predictions of the isotropic theory of elasticity, atoms' displacements in different crystallographic directions differ significantly, both for α-Fe and W. For directions of type <100>, the displacements near the voids are positive. In α-Fe, as shown in Figure 3, a change in the sign of the displacement to negative occurs for atoms located from the void center by a distance of more than 42 Å (more than 12 lattice parameters). The maximum atoms' displacements from the site of the ideal lattice are observed in the direction <111>. It should be noted that according to the simulation results for voids of different sizes (up to 20 Å [16]), the dependences have a similar form.

From the analysis of the results it follows that the trace of the strain tensor is not zero, and the equations for vacancy fluxes, and consequently the kinetic equations for the void growth rate must contain additional terms due to the elastic field:

$$\nabla Sp\varepsilon \neq 0. \qquad (8)$$

Therefore, in the next section, we present the derivation of equations for the shifting rate of void surface elements based on the expressions for the vacancy fluxes that we obtained earlier in [10].

## 4. Shifting rate of the void surface elements for the different crystallographic directions

It can be shown [17] that the movement of the shifting rate of void surface element in a certain direction is determined by the equation:

$$\frac{dR}{dt} = -\left(\vec{n}, \vec{j}\right), \qquad (9)$$

where $R$ is the radius of the void, $\boldsymbol{n}$ is the normal to the void surface, $\boldsymbol{j}$ is vacancy flux density to the void surface.

To evaluate the influence of the elastic field on the vacancy flux in the region near the voids and obtain analytical solutions, the method of successive approximations is used. And as a first approximation, similar to [18], we choose the solution of the diffusion equation for the vacancy concentration in that the influence of the field is not taken into account. Then the change rate of the void radius $R$ is described by equation [18]:

$$\frac{dR}{dt} = C_{eq} D_V \left[ \frac{R_G}{R_G - R} R^{-1} \left( \Delta + 1 - \exp\left(\frac{2\gamma V^f}{kTR}\right) \right) \right], \tag{10}$$

where $c_{eq}$ is the equilibrium vacancy concentration for the flat surface, $V^f$ is the vacancy formation volume, $\Delta = \frac{c_m - c_{eq}}{c_{eq}}$ is the oversaturation of vacancies, $c_m = c(R_G)$, where $R_G$ is the average distance between voids.

Transformations for getting the equations for the shifting rate of the void surface elements for various crystallographic directions taking into account the field of elastic strains are given in [9]. Here, we restrict ourselves to providing final equations for the shifting rate of the void surface elements:

$$\frac{dR_{100}}{dt} = C_{eq} D_V \left[ \frac{R_G}{R_G - R} R^{-1} \left( \Delta + 1 - \exp\left(\frac{2\gamma V^f}{kTR}\right) \right) - \frac{K^V}{kT} \exp\left(\frac{2\gamma V^f}{kTR}\right) \frac{\partial Sp\varepsilon}{\partial x} \right], \tag{11}$$

$$\frac{dR_{110}}{dt} = \frac{C_{eq} D_V}{\sqrt{2}} \left[ \frac{\sqrt{2} R_G}{R_G - R} R^{-1} \left( \Delta + 1 - \exp\left(\frac{2\gamma V^f}{kTR}\right) \right) - \frac{K^V}{kT} \exp\left(\frac{2\gamma V^f}{kTR}\right) \left( \frac{\partial Sp\varepsilon}{\partial x} + \frac{\partial Sp\varepsilon}{\partial y} \right) \right], \tag{12}$$

$$\frac{dR_{111}}{dt} = \frac{C_{eq} D_V}{\sqrt{3}} \left[ \frac{\sqrt{3} R_G}{R_G - R} R^{-1} \left( \Delta + 1 - \exp\left(\frac{2\gamma V^f}{kTR}\right) \right) - \frac{K^V}{kT} \exp\left(\frac{2\gamma V^f}{kTR}\right) \left( \frac{\partial Sp\varepsilon}{\partial x} + \frac{\partial Sp\varepsilon}{\partial y} + \frac{\partial Sp\varepsilon}{\partial z} \right) \right]. \tag{13}$$

The components of the strain tensor for each of the crystallographic directions, the trace of the strain tensor, and its derivatives to the corresponding coordinates are calculated based on the approximated simulation results of atom displacements near the void surface. After that we can calculate the shifting rate of the void surface elements using equations (11-13). It was done for various oversaturation's in a wide temperature range. Figures 5, 6 show the normalized dependences of the mentioned void growth rates for α-Fe and W at the oversaturation $\Delta = 20$.

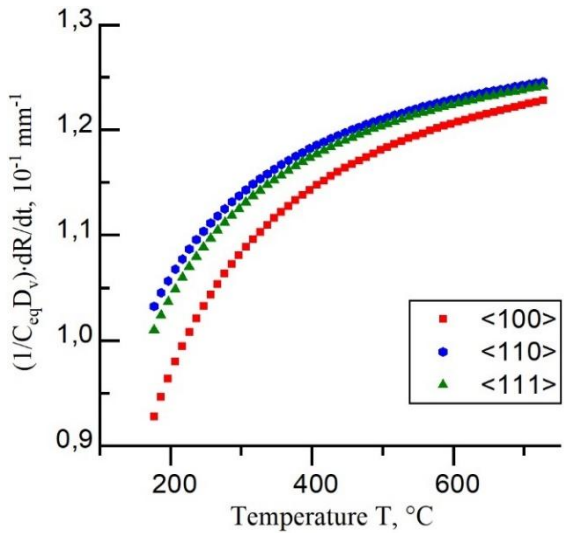
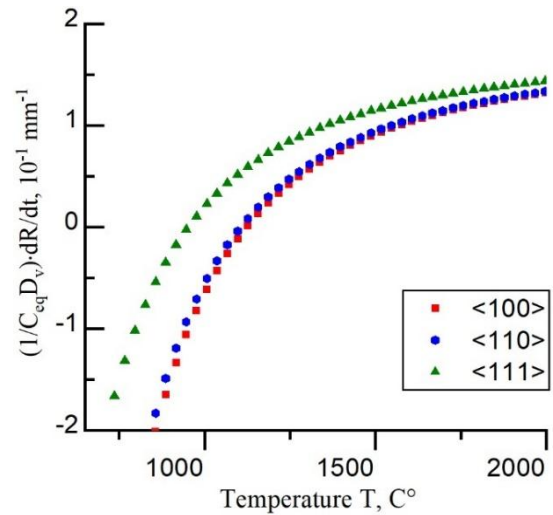

**Fig. 5.** Shifting rate of the void surface elements for various crystallographic directions in α-Fe (R = 15.14 Å)

**Fig. 6.** Shifting rate of the void surface elements for various crystallographic directions in W (R = 10.96 Å)

In α-Fe void growth rate are positive for all crystallographic directions, however, the void growth rate significantly slows down in the direction of the <100> type due to the negative contribution of the elastic field. For directions of the <110> and <111> types the growth rates are quite close. In W for the <111> direction the elastic displacement field makes a significant contribution to the shifting rate of the void surface element.

## 5. Discussion

As can be seen from equations (11-13), the additional term due to the elastic displacement field can contributes to the equations for the shifting rate of the void surface elements. According to the simulation results atomic displacements for α-Fe and W in the <100> direction are positive, and the corresponding contribution of the elastic field to shifting rate in equation (11) is negative. For this direction the void growth rate is slowed down, while the contributions in the directions <110> and <111> (see equations (12) and (13)) are positive.

In α-Fe, the shifting rate in the direction of the <100> type differs significantly from the directions of the <110>, <111> type. In W the atoms' displacements in the <100> direction also are positive, however displacements at void surface in W are notably less than in Fe, and then influence of elastic field to shifting rate of void surface element is lower.

Thus, for α-Fe and W, one of the reasons that influence the change in the shape of the initially spherical voids is the vacancy flux anisotropy due to the features of the atomic structure in the vicinity of the void. The asymmetry of atomic displacements in the vicinity of voids in α Fe and W slows down the shifting rate of the void surface elements in directions that coincide or are close to the crystallographic type <100>. The results show that initially the spherically symmetric void ccould change its shape, and due to the slowing down of the shifting rate of the void surface elements along the <100> directions, the void shape will gradually be faceted by {100} planes.

## 6. Conclusion

The effect of the anisotropy of atoms' displacements in the vicinity of nanovoids in α-Fe and W on the shifting rate of the void surface elements is studied. Using the Advanced Molecular Statics method, the atomic structure is simulated in the vicinity of voids in α-Fe and W. In contrast to the predictions of the isotropic theory of elasticity, the atoms' displacements in different crystallographic directions differ significantly, both for α-Fe and W. Should be emphasized that for directions of the type <100> displacements near the void are positive. Based on the simulation results for atoms' displacements, the dependences of the shifting rate of the void surface elements are calculated for various oversaturation's in a wide temperature range. Asymmetry of atoms' displacements in the vicinity of voids leads to a slowdown in the movement of surface elements in directions that are close to the <100>, that leads to a change in the void of the initially spherical shape to a shape faceted by {100} planes.

**References**

[1]  Was G S Fundamentals of radiation materials science (New York: Springer) p 827.
[2]  Garner F 2012 Compr. Nucl. Mater. 4 33-95
[3]  Chen C 1973 Phys. Status Solidi. 16 197-210
[4]  Zinkle S 2012 Compr. Nucl. Mater. 1 65-98.
[5]  Osetsky Y and Bacon D 2010 Philos. Mag. 90 945-61
[6]  Niwase K, Ezawa T, Fujita F, Kusanagi H and Takaku H 1988 Radiat. Eff. 106 65-76
[7]  Han G, Wang H, Lin D, Zhu X, Hu S and Song H 2017 Comput. Mater. Sci. 133 22-34
[8]  Mikheev A, Nazarov A, Ershova I and Zaluzhnyi A 2015 Defect Diffus. Forum 363 91-98
[9]  Nazarov A, Mikheev A and Melnikov A 2018 Preprint arXiv:1811.01422, 2020 J. Nucl. Mat. (DOI: 10.1016/j.jnucmat.2020.152067)
[10] Nazarov A and Mikheev A 2015 Defect Diffus. Forum 363 112-119
[11] Eshelby J 1956 Sol. State Phys. 3 79–144

[12] Valikova I, Nazarov A and Mikheev A 2006 Defect Diffus. Forum 249 55-60
[13] Valikova I and Nazarov A 2008 Defect Diffus. Forum 277 125-32
[14] Valikova I and Nazarov A 2008 Phys. Met. Metallogr. 105 544-52
[15] Valikova I and Nazarov A 2010 Phys. Met. Metallogr. 109 220-26
[16] Nazarov A, Ershova I and Volodin Y 2018 KnE Mater. Sci. 2018 451-57
[17] Nazarov A Phys. Met. Metallogr. 1973 35 №3 184–186
[18] Cheremskoy P G, Slyozov V V and Betehtin V I 1990 Voids in Solid State (Moscow: Energoatomizdat) p 376 (in Russian)